\documentclass{article}

\usepackage{PRIMEarxiv}

\usepackage[utf8]{inputenc} 
\usepackage[T1]{fontenc}    
\usepackage{hyperref}       
\usepackage{url}            
\usepackage{booktabs}       
\usepackage{amsfonts}       
\usepackage{nicefrac}       
\usepackage{microtype}      
\usepackage{lipsum}
\usepackage{fancyhdr}       
\usepackage{graphicx}       
\graphicspath{{media/}}     
\usepackage{amsmath}
\usepackage{natbib}

\title{Electoral Intelligence: A Neural Zero-Inflated Negative Binomial Model for Forecasting Swedish Municipal Elections
}

\author{
  Haodong Qi \\
  Department of Global Political Studies \\
  Malmö University \\
  Malmö\\
  \texttt{Haodong.qi@mau.se} \\
  \AND
  Stefano M. Iacus \\
  Institute for Quantitative Social Science \\
  Havard Univerity \\
    \AND
  Pieter Bevelander \\
  Department of Global Political Studies \\
  Malmö University \\
    \AND
  Derek S. Hutcheson \\
  Department of Global Political Studies \\
  Malmö University \\
}

\begin{document}
\maketitle

\begin{abstract}
Forecasting local elections is difficult because electoral competition is geographically heterogeneous, highly overdispersed, and characterized by many party–municipality combinations with little or no electoral activity. This article develops a bounded neural zero-inflated negative binomial (ZINB) model for forecasting party-level vote counts in Swedish municipal elections. The model combines a count component for expected votes, a zero-inflation component for structural electoral absence, municipality-specific dispersion, and learned embeddings capturing latent party, municipality, and party-municipality structure. Model training relies on data from the 2014, 2018, and 2022 Swedish municipal elections and performance is evaluated on a held-out set of municipalities. Testing results show strong predictive accuracy in both vote counts and vote shares. Explanatory diagnostics indicate that latent party structure, local party-municipality relationships, electoral persistence, organizational capacity, and candidate-electorate representational similarity are among the most important determinants of predictive performance. Following model evaluation and diagnostics, the selected specification is refit on all historical data and applied to generate forecasts for the 2026 municipal elections. The proposed framework provides a flexible approach for studying local electoral competition in multiparty systems characterized by heterogeneous party presence and substantial geographic variation.
\end{abstract}

\section{Introduction}
Election forecasting occupies a central position in empirical political science because it provides both a practical tool for anticipating electoral outcomes and a means of evaluating theories of political behavior. A long tradition of forecasting research has demonstrated that election results can often be predicted well in advance using political and economic fundamentals, historical voting patterns, opinion polls, or combinations thereof. Beyond prediction itself, successful forecasts can reveal the underlying structure of electoral competition and improve our understanding of how voters respond to political, economic, and institutional conditions \citep{LewisBeckTien2015,LewisBeckEtAl2025}. 

Despite considerable methodological progress, most election forecasting models have been developed for national elections and often focus on two-party systems (such as the US election). Forecasting in multiparty settings presents additional challenges because party support is inherently compositional and electoral competition is distributed across multiple parties simultaneously. Recognizing these difficulties, recent work has extended forecasting methods to multiparty systems, primarily through Bayesian approaches that combine polling data with structural fundamentals \citep{StoetzerEtAl2019}. However, much of this literature remains oriented toward national elections and relies heavily on polling information, which are typically unavailable or sparse in local elections. 

Swedish municipal elections provide a particularly demanding forecasting environment. Electoral competition involves multiple national parties, hundreds of local parties, and substantial geographic heterogeneity. These characteristics lead to a large number of party--municipality combinations in which parties receive either very few votes or no votes at all. The resulting data exhibit two features that challenge conventional forecasting models. First, party support contains many excess zeros arising from limited local party presence. Second, electoral support varies considerably across municipalities, producing substantial over-dispersion relative to standard count models. Traditional linear forecasting models are not well suited to accommodate these characteristics simultaneously. 

At the same time, local elections contain rich sources of information that are often ignored in conventional electoral forecasting. Historical electoral performance, local candidate-list characteristics, municipality-specific voter composition, and the similarity between candidates and electorates may all contribute to electoral success. Furthermore, local electoral competition is shaped by persistent party identities and municipality-specific political contexts that are difficult to represent using standard fixed-effect approaches. These features suggest the need for forecasting models that are flexible enough to capture nonlinear relationships while remaining interpretable and capable of quantifying uncertainty. 

This paper develops a bounded neural zero-inflated negative binomial (ZINB) framework for forecasting Swedish municipal elections. The model combines several components. First, a zero-inflated negative binomial likelihood accommodates excess zeros and over-dispersion. Second, the expected vote count is bounded by the number of eligible voters, ensuring substantively plausible predictions. Third, party, municipality, and party--municipality interaction embeddings allow the model to learn latent political structure across local electoral environments. Finally, municipality-specific dispersion parameters allow electoral uncertainty itself to vary across locations. 

Methodologically, the paper contributes to a growing literature that integrates modern machine-learning techniques with established electoral forecasting principles. Whereas most existing multiparty forecasting models focus on national elections and polling aggregation, the proposed approach emphasizes local electoral forecasting using electoral records, candidate information, and voter composition data. These information are publicly available on the official Swedish electoral homepage (\url{www.val.se}), and updated multiple times during the year preceding to the election, allowing timely forecast revisions. 

The framework therefore ingests timely information, and combines the predictive flexibility of neural networks with the probabilistic structure and interpretability of count-data models. Substantively, the paper provides a forecast of the 2026 Swedish municipal elections and offers several diagnostic perspectives on local electoral competition. Beyond point forecasts, the model estimates the probability that parties receive meaningful electoral support, predicted vote counts, vote shares, and forecast uncertainty. The results provide insight not only into expected electoral outcomes but also into the geographic distribution of electoral presence, party strongholds, and the variation in forecast reliability. 

This paper presents forecasts for the 2026 Swedish municipal elections using information available as of August 28, 2026. The forecasts are generated prior to election day (on September 13, 2026) and therefore constitute genuinely out-of-sample predictions. Future versions of the paper will evaluate forecast accuracy against the realized election results which will be published.

\section{Statistical Model and Implementation}
\label{sec:model}
The unit of analysis is a party--municipality--year observation. Let \(y_{pmt}\) denote the number of votes received by party \(p\) in municipality \(m\) at election year \(t\), and let \(N_{mt}\) denote the number of eligible voters in the corresponding municipality-year. The goal is to model and forecast the conditional distribution of \(y_{pmt}\) using observed predictors, party and municipality latent structure, and historical electoral information. Municipal election data have several important features. First, the outcome is a non-negative count. Second, many party--municipality observations are zeros, especially for parties that have weak or no local presence. Third, vote counts are strongly overdispersed relative to a Poisson model. Given these data characteristics, we use a zero-inflated negative binomial model with a neural-network parameterization of the mean, zero-inflation probability, and dispersion. 

\subsection{Zero-inflated negative binomial likelihood} 
For each party--municipality--year observation \(i=(p,m,t)\), the model assumes 
\begin{align}
     y_i \sim \mathrm{ZINB}(\mu_i,\alpha_i,\pi_i), 
\end{align}
where \(\mu_i>0\) is the negative-binomial mean, \(\alpha_i>0\) is the dispersion parameter, and \(\pi_i \in (0,1)\) is the probability that the observation belongs to the structural-zero component. The negative binomial component is parameterized as 
\begin{align}\label{eq:nb_mu}
    \mathbb{E}(y_i \mid \mu_i,\alpha_i)=\mu_i, \\
\end{align}

and 
\begin{align}\label{eq:nb_var}
       \mathrm{Var}(y_i \mid \mu_i,\alpha_i) = \mu_i+\alpha_i\mu_i^2.
\end{align}
 
Thus \(\alpha_i\) measures over-dispersion relative to the Poisson model. When \(\alpha_i\) is close to zero, the variance approaches the Poisson variance. Larger values of \(\alpha_i\) indicate greater residual volatility around the predicted mean. The zero-inflated likelihood is \[ \Pr(y_i=0) = \pi_i + (1-\pi_i)\Pr_{\mathrm{NB}}(0\mid \mu_i,\alpha_i), \] and for positive counts, \[ \Pr(y_i=y) = (1-\pi_i)\Pr_{\mathrm{NB}}(y\mid \mu_i,\alpha_i), \qquad y>0. \] 
Therefore, the log-likelihood contribution is 
\begin{align}\label{eq:lik}
    \ell_i = \mathbb{I}\{y_i=0\} \log\left[ \pi_i + (1-\pi_i)\Pr_{\mathrm{NB}}(0\mid \mu_i,\alpha_i) \right] + \mathbb{I}\{y_i>0\} \left[ \log(1-\pi_i) + \log \Pr_{\mathrm{NB}}(y_i\mid \mu_i,\alpha_i) \right].
\end{align}

\subsection{Bounded count mean} 
The count component predicts the expected number of votes conditional on the observation not being a structural zero. Instead of using an unbounded log-link for the mean, we model the expected vote count as eligible voters multiplied by a predicted party vote share. Let \(\eta_i^\mu\) denote the latent count index produced by the neural count tower. The mean is defined as 
\begin{align}\label{eq:mu}
    \mu_i = N_{mt}\mathbb{S}(\eta_i^\mu), 
\end{align}

where \(\mathbb{S}(\eta_i^\mu)=1/(1+\exp(-\eta_i^\mu))\) is the sigmoid function. Since \(0<\mathbb{S}(\eta_i^\mu)<1\), this specification ensures \[ 0 < \mu_i < N_{mt} \] which is substantively important because a single party cannot receive more votes than there are eligible voters in a municipality-year. 

In the implementation, the model works with \(\log \mu_i\): \[ \log \mu_i = \log N_{mt} + \log \mathbb{S}(\eta_i^\mu). \] The offset \(\log N_{mt}\) is retained, thus the neural network effectively predicts a bounded vote share rather than an unbounded count rate. 

\subsection{Count tower} 
Let \(x_i\) denote the vector of numeric predictors entering the count component. These may include, for example, previous electoral performance, candidates' characteristics, and voters' compositional characteristics. The model first applies feature-specific gates: \[ \widetilde{x}_i = x_i \odot g^\mu, \] where \(\odot\) denotes element-wise multiplication and \[ g^\mu_j = \mathbb{S}(\gamma^\mu_j). \] 
where, \(\gamma^\mu_j\) is the raw gate parameter for predictor \(j\) in the count component. \(\mathbb{S}\) is the sigmoid function forcing \( g^\mu_j \in (0,1) \).

The model includes learned embeddings for parties and municipalities. Let \(\mathbf{e}^P_p\) denote the party embedding and \(\mathbf{e}^M_m\) denote the municipality embedding. The main implementation also permits a party--municipality interaction embedding \(\mathbf{e}^{PM}_{pm}\), which captures persistent local affinities between specific parties and municipalities. The count network is then 
\[ h_i^\mu = f_\mu \left( \widetilde{x}_i, \mathbf{e}^P_p, \mathbf{e}^M_m, \mathbf{e}^{PM}_{pm} \right), \] 
where \(f_\mu(\cdot)\) is a feed-forward neural network with nonlinear activation functions and dropout. 

The final layer of this network produces the latent vote-share index \[ \eta_i^\mu = w_\mu^\top h_i^\mu + b_\mu. \] This index is then mapped into Eq.~\ref{eq:mu} to obtain the negative negative-binomial mean of vote counts.

\subsection{Zero-inflation tower} 
The zero-inflation component models whether a party--municipality observation belongs to the structural-zero state. Let \(z_i\) denote the vector of numeric predictors entering the zero-inflation component, and let \(b_i\) denote binary predictors such as whether a party had any votes in previous elections or candidate presence. The numeric zero-inflation predictors are passed through a separate gate: 
\[ \widetilde{z}_i = z_i \odot g^\pi, \] 
where \[ g^\pi_j = \mathbb{S}(\gamma^\pi_j). \] 

The zero-inflation network is 
\[ h_i^\pi = f_\pi \left( \widetilde{z}_i, b_i, \mathbf{e}^P_p, \mathbf{e}^M_m, \mathbf{e}^{PM}_{pm} \right). \] 

The output is a logit for the structural-zero probability: 
\[ \eta_i^\pi = w_\pi^\top h_i^\pi + b_\pi, \] and therefore \[ \pi_i = \mathbb{S}(\eta_i^\pi). \] A high value of \(\pi_i\) means that the model assigns high probability to the party being structurally absent or electorally noncompetitive in the municipality--year units. 

\subsection{Dispersion component} 
The negative-binomial dispersion parameter controls residual variation in vote counts after conditioning on the mean. In the main specification, dispersion is municipality-specific. The model uses the municipality embedding to produce a dispersion index: 
\[ \eta_m^\alpha = f_\alpha(\mathbf{e}^M_m). \] 

The output is clipped for numerical stability and transformed using the softplus function: \[ \alpha_m = \mathrm{softplus}(\eta_m^\alpha) = \log(1+\exp(\eta_m^\alpha)). \] This guarantees \(\alpha_m>0\). Substantively, larger values of \(\alpha_m\) indicate municipalities with greater residual electoral volatility. In essence, the dispersion component can be specified as row-specific (i.e., \(\alpha_i\) vary by party--municipality--year) or global \(\alpha\). However, allowing row-specific \(\alpha_i\) introduces unnecessary noise which may induce overfitting, whilst global \(\alpha\) is too restrictive that might potentially lead to underfitting. To balance, we assume municipality-specific dispersion \(\alpha_m\). This choice allows us to capture municipal heterogeneity of electoral uncertainty which is central to local election forecasting. 

\subsection{Shrinkage gates} The count and zero-inflation towers each contain elastic-net shrinkage gates. In the count tower, the raw gate vector \(\gamma^\mu\) is penalized by 
\[ \Omega_\mu = \lambda_\mu \left[ \rho \sum_j |\gamma^\mu_j| + (1-\rho)\sum_j (\gamma^\mu_j)^2 \right]. \] 

Similarly, in the zero-inflation tower, \[ \Omega_\pi = \lambda_\pi \left[ \rho \sum_j |\gamma^\pi_j| + (1-\rho)\sum_j (\gamma^\pi_j)^2 \right]. \] 

The tuning parameters \(\lambda_\mu\) and \(\lambda_\pi\) control the amount of shrinkage in the count and zero-inflation components, respectively. In the empirical implementation, \(\rho=0.5\), giving equal weight to the \(L_1\) and \(L_2\) parts of the elastic-net penalty. Because the penalty is applied to the raw gate logits \(\gamma_j\), regularization shrinks gate logits toward zero. Note, since  $\mathbb{S}(0)=0.5$, a gate value of \(0.5\) represents the neutral scaling point. Gates above \(0.5\) indicate features that the model moves above the neutral baseline, while gates below \(0.5\) indicate features that are attenuated relative to that baseline. In the main forecasting specification, the gates are used as continuous shrinkage weights rather than as a hard variable-selection rule. 

\subsection{Objective function} 
The model is estimated by minimizing the regularized negative log-likelihood: 
\begin{align}\label{eq:loglikfun}
    \mathcal{L} = -\frac{1}{n}\sum_{i=1}^{n}\ell_i + \Omega_\mu + \Omega_\pi + \Omega_{\mathrm{emb}},
\end{align}

where \(\ell_i\) is the zero-inflated negative binomial log-likelihood contribution as in Eq.~\ref{eq:lik}, \(\Omega_\mu\) and \(\Omega_\pi\) are the elastic-net gate penalties, and \(\Omega_{\mathrm{emb}}\) denotes \(L_2\) penalties on the embedding parameters. 

The model also allows the zero component of the likelihood to be weighted by a scalar \(w_0\), although in the main specification this scalar is fixed at one. The full model therefore jointly learns $\mu_i,\alpha_m,\pi_i$  for each party--municipality--year observation. The final point forecast is the unconditional expected number of votes: \[ \widehat{y}_i = (1-\widehat{\pi}_i)\widehat{\mu}_i. \] 

For forecast interpretation, we transform predicted support as a share of eligible voters: 
\begin{align}\label{eq:pred_share}
    \widehat{s}_i = \frac{\widehat{y}_i}{N_{mt}}.
\end{align}

\subsection{Workflow}
\begin{figure}[!ht]
    \centering
    \includegraphics[width=.8\linewidth]{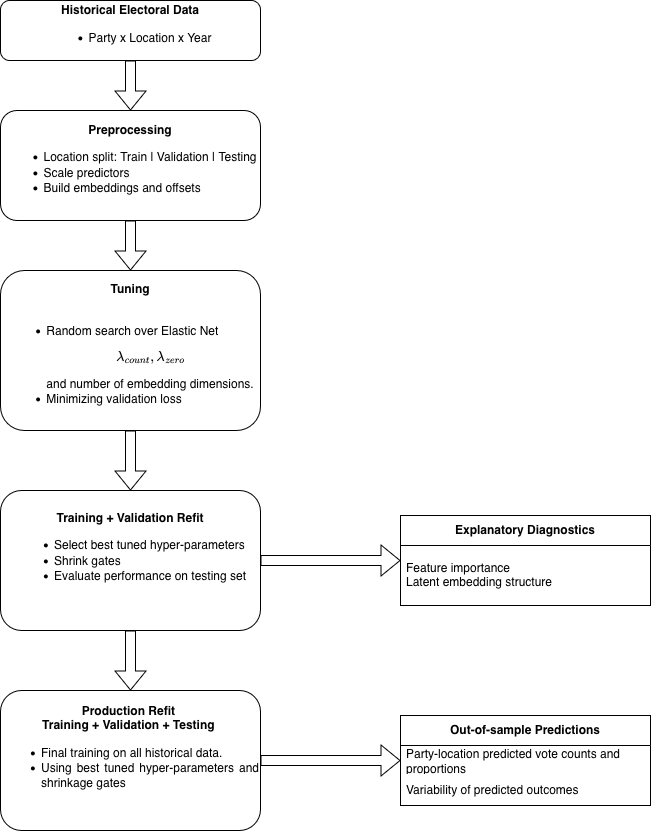}
     \caption{Workflow from historical data to the 2026 production forecast. The historical electoral panel is first preprocessed and split by municipality into training, validation, and testing sets. Hyperparameters are selected using a coarse-to-fine search over elastic-net gate penalties and embedding dimensions. The selected model is refit on the training and validation data for held-out test evaluation and explanatory diagnostics. The final production model is then refit on all historical data and used to generate 2026 party--municipality forecasts.} 
    \label{fig:workflow}
\end{figure}

Figure~\ref{fig:workflow} summarizes the full estimation and forecasting workflow. The key distinction is between the train--validation refit, which is used for model evaluation and explanation, and the production refit, which uses all historical data to generate the final 2026 forecasts. The starting point is a historical electoral panel in which each observation corresponds to a party--location--year tuple. Let \(y_{pmt}\) denote the number of votes received by party \(p\) in municipality \(m\) at election year \(t\). The historical data are used to estimate and evaluate the model, while the 2026 observations are held out for out-of-sample forecasting.

First, the historical panel is preprocessed. Municipalities are split into training, validation, and testing sets so that the evaluation is location-based rather than based on randomly sampled rows. This prevents observations from the same municipality appearing simultaneously in both training and testing data. Numeric predictors are standardized, and the learned party, municipality, and party--municipality indices are constructed for the embedding layers. 

Second, hyper-parameters are selected using a two-stage coarse-to-fine procedure. In the first stage, candidate values for \(\lambda_\mu\), \(\lambda_\pi\), and the embedding dimensions (for parties, municipalities, and party--municipality interactions) are sampled over prespecified grids or log-uniform ranges. Each candidate model is trained on the municipalities in the training set and evaluated on the validation municipalities. In the second stage, the best candidates from the first stage are locally refined by searching over multiplicative neighborhoods of their penalty values while holding the embedding dimensions fixed. The validation objective is to minimize the negative log-likelihood function Eq.~\ref{eq:loglikfun}.

Third, after selecting the best hyper-parameters, the model is refit on the combined training and validation data and evaluate predictive performance on the held-out test municipalities. The same train--validation refit model is also used for explanatory diagnostics, including ablation analysis, embedding structure, and municipality-specific dispersion estimates.

Fourth, the final production model is fit on all available historical data, combining the training, validation, and testing sets. This production refit uses the selected hyper-parameters and the same shrinkage-gated zero-inflated negative binomial architecture. In the main specification, all predictors are retained and the elastic-net gates are used as continuous shrinkage devices rather than as hard feature-selection rules. The count component uses a bounded vote-share mean as in Eq.~\ref{eq:mu}.

Finally, the production model is applied to the 2026 party--municipality observations to generate out-of-sample forecasts. For each party--municipality pair, the model produces the expected count
\[
\widehat{y}_{pm,2026}
=
(1-\widehat{\pi}_{pm,2026})\widehat{\mu}_{pm,2026},
\]
where \(\widehat{\pi}_{pm,2026}\) is the predicted structural-zero probability and \(\widehat{\mu}_{pm,2026}\) is the predicted negative-binomial mean. Forecasts are reported as predicted shares of eligible voters,
\[
\widehat{s}_{pm,2026}
=
\frac{\widehat{y}_{pm,2026}}{N_{m,2026}}.
\]
These predictions form the basis for the 2026 forecast maps and subsequent uncertainty diagnostics.

\section{Swedish Electoral Data} \label{sec:data} 
The empirical analysis combines electoral results, candidate-list information, and demographic characteristics of the electorate at the municipality level. Historical data are obtained from the Swedish Election Authority (Valmyndigheten), which provides open election data covering municipal elections, candidate registrations, party participation, and electorate statistics\footnote{\url{https://www.val.se/valresultat-och-statistik/statistik-och-data/radata-fran-val-2002-2022}}. The estimation sample covers the municipal elections of 2014, 2018, and 2022. The forecast sample corresponds to the 2026 municipal election. Historical election outcomes are used exclusively for model estimation, validation, and testing, whereas 2026 observations are reserved for forecasting purposes. 

\subsection{Election Data} The core electoral dataset records votes received by each party in each municipality. In addition to vote totals, the data contain information on valid ballots, invalid ballots, turnout, and the number of eligible voters. Because Swedish municipal elections feature both national and local parties, the resulting dataset contains a large number of party--municipality combinations with very limited electoral activity or no votes at all. This sparsity motivates the zero-inflated modeling strategy introduced in Section~\ref{sec:model}. To facilitate the embedding structure of the neural model, each party and municipality is assigned a unique numerical identifier. An additional identifier is created for each observed party--municipality combination, allowing the model to learn persistent local affinities between particular parties and municipalities. 

\subsection{Electorate Composition} The Election Authority publishes municipality-level tabulations of eligible voters by sex, citizenship status, age group, and first-time voter status. These tabulations are transformed into proportions of the eligible electorate. The resulting variables measure the shares of women, Swedish citizens, first-time voters, and four age groups (18--29, 30--49, 50--64, and 65 years or older) within each municipality. Because these variables describe the electorate rather than individual parties, they vary across municipalities and election years but are common to all parties competing within the same municipality-election. 

\subsection{Candidate Characteristics} The electoral data are supplemented with information on party candidate lists. For each party--municipality observation, candidate-level records are aggregated into summary measures describing the demographic composition of the list. These measures include the proportion of women candidates, the age composition of candidates, candidate-list size, and indicators for incomplete age information. The age categories correspond to those used in the electorate data, ensuring direct comparability between candidates and voters. Candidate characteristics are particularly relevant in municipal elections because local candidate recruitment often differs considerably across municipalities, even within the same party organization. 

\subsection{Candidate--Electorate Similarity} To capture descriptive representation, we construct similarity measures comparing the composition of party candidate lists with the composition of the local electorate. Age similarity is measured using the Bhattacharyya coefficient\footnote{The Bhattacharyya coefficient is a statistical measure used to calculate the amount of overlap or similarity between two probability distributions or samples.}, which compares the candidate age distribution with the electorate age distribution. The coefficient ranges between zero and one, with larger values indicating greater correspondence between the age structure of candidates and voters. Gender similarity is calculated as one minus the absolute difference between the share of women among candidates and the share of women among eligible voters. Values close to one indicate that the candidate list closely mirrors the gender composition of the electorate. These measures are intended to capture whether parties field candidate lists that resemble the populations they seek to represent. 

\subsection{Lagged Electoral Performance} Electoral persistence is incorporated through lagged election outcomes. For each party and municipality, the dataset records the vote total obtained in the previous municipal election whenever such information is available. In addition to raw lagged vote totals, the data include a logarithmic transformation of previous votes and indicators identifying observations for which no previous electoral result exists. This distinction is important because the absence of a lagged observation may reflect party entry, party exit, or local organizational absence rather than weak electoral support. 

\subsection{Forecast Data for 2026} 
Forecast covariates for the 2026 election are compiled from the Election Authority's preliminary election files. These files contain candidate registrations, participating parties, and updated municipality-level counts of eligible voters. At the time of data collection (4 July 2026), detailed demographic tabulations of the 2026 electorate were not yet available. Consequently, municipality-level electorate composition variables are carried forward from the corresponding 2022 values. Candidate information, party participation, and eligible-voter counts are obtained directly from the 2026 election files. No 2026 vote outcomes are used during model estimation, hyperparameter tuning, explanatory analysis, or model evaluation. The 2026 observations are reserved exclusively for forecasting.

\section{Model Performance and Explanatory Insights} \label{sec:results} 

\subsection{Predictive Performance} 
\begin{figure}[!ht]
    \centering
    \includegraphics[width=1\linewidth]{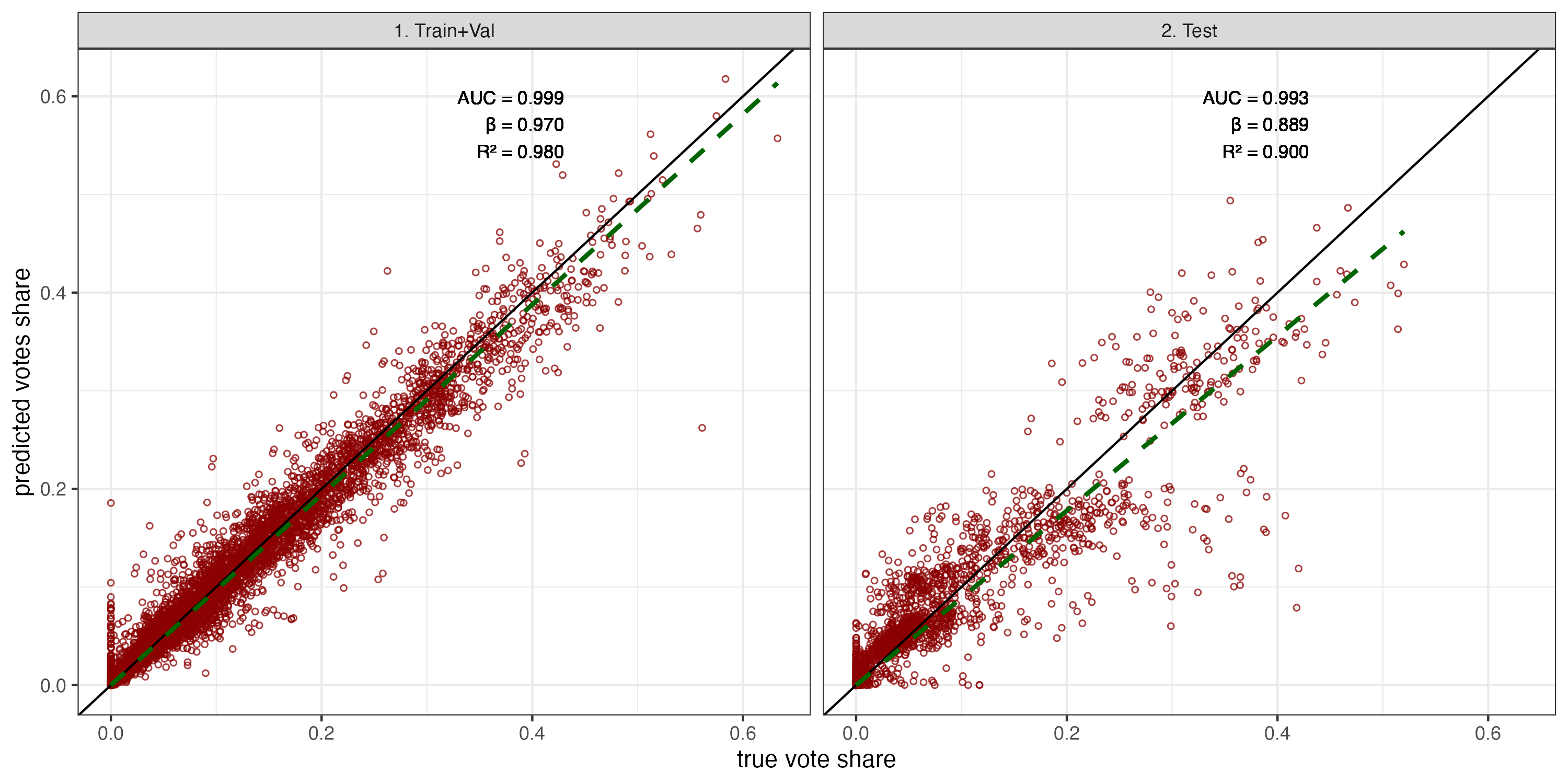}
    \caption{Observed and predicted party support among eligible voters.}
    \label{fig:accuracy}
\end{figure}

Figure~\ref{fig:accuracy} summarizes predictive performance on the training--validation and held-out test samples. Performance is evaluated using party support among eligible voters, \[ s_{pmt} = \frac{y_{pmt}}{N_{mt}}, \] and its corresponding prediction, \[ \widehat{s}_{pmt} = \frac{\widehat{y}_{pmt}}{N_{mt}}, \] where \(N_{mt}\) denotes the number of eligible voters in municipality \(m\) during election year \(t\). 

The left panel of Figure~\ref{fig:accuracy} shows predictions from the model refit on the combined training and validation data. Predicted support closely follows observed support throughout the range of observed outcomes. The coefficient of determination is \(R^2=0.98\), indicating that the model explains nearly all variation in observed party support within the estimation sample. The estimated calibration slope is \(\beta=0.97\), suggesting only limited deviation from perfect calibration. 

The right panel reports out-of-sample performance on municipalities not used during model training or hyperparameter tuning. Although predictive accuracy declines relative to the estimation sample, performance remains strong. The model attains an out-of-sample \(R^2=0.9\), while the calibration slope remains high at \(\beta=0.89\). The area under the receiver operating characteristic curve is near 1, indicating excellent separation between zero and non-zero electoral outcomes. The test results therefore suggest that the model generalizes well to previously unseen municipalities. 

Several features of the test-panel scatterplot are noteworthy. First, the dispersion of prediction errors increases for larger parties and stronger party-municipality combinations. This pattern is expected because vote counts become more heterogeneous as support increases. Second, the fitted relationship is slightly flatter than the 45-degree reference line, indicating mild underprediction at the upper end of the support distribution. Nevertheless, the overall fit remains close to the identity line across most observed values. The results overall suggest that the bounded neural zero-inflated negative binomial model captures both the extensive margin of local electoral competition (whether parties receive support at all) and the intensive margin of electoral strength once parties are electorally present.

\subsection{Feature Importance} 
Although the primary purpose of the model is forecasting, the ablation analysis provide useful insights into the information used by the model when generating predictions. Table~\ref{tab:ablation_elasticity} reports  an ablation measure based on the proportional increase in mean absolute error after removing a predictor from the fitted model. 

The largest ablation effects are associated with the learned embeddings. Removing the party embedding increases prediction error by almost ninety-fold, while removing the party--municipality interaction embedding increases prediction error by more than six-fold. These results indicate that persistent differences among parties and local party--municipality relationships contain substantially more predictive information than any individual observed covariate. The municipality embedding is also important, suggesting that local political environments contain systematic patterns that are only partially captured by observed demographic and organizational characteristics. 

Among the observed predictors, historical electoral performance remains one of the strongest sources of predictive information. However, the results reveal an important distinction between raw and logarithmic measures of past support. Removing the logarithm of lagged votes increases the mean absolute error by approximately 150\%, whereas removing the raw lagged vote count increases error by only about 19\%. This pattern suggests that proportional differences in previous electoral support are more informative than absolute vote totals. The logarithmic transformation compresses large values and emphasizes relative variation among parties and municipalities, making it a more natural predictor in a model whose ultimate target is support relative to the size of the electorate rather than unrestricted vote counts. 

Candidate-list characteristics also contribute substantially to predictive performance. In particular, the number of candidates is among the strongest observed predictors, with an ablation effect comparable to that of lagged electoral performance. A plausible interpretation is that candidate-list size reflects local organizational capacity. Parties capable of recruiting larger candidate slates are typically those with stronger activist networks, greater campaign presence, and a more established local organization. Candidate-list size therefore acts as a proxy for local party strength and electoral competitiveness. 

The composition of candidate lists likewise appears informative. The age distribution of candidates, especially the shares of candidates aged 30--49 and 50--64, produces noticeable increases in forecasting error when removed from the model. These age groups correspond closely to the core working-age population that dominates much of local political participation and party recruitment. Candidate age composition may therefore capture organizational experience, leadership resources, and the ability of parties to mobilize electorally active segments of the population. Candidate gender composition contributes less predictive information than age composition but nevertheless improves forecast accuracy, indicating that patterns of candidate recruitment contain information beyond simple party identity. 

Particularly noteworthy are the similarity measures comparing candidate composition with electorate composition. The age-similarity measure is among the most influential observed predictors in the model, generating a larger deterioration in predictive performance than most candidate-level or voter-level demographic variables. One interpretation is that parties whose candidate lists more closely resemble the age structure of the local electorate are better positioned to attract support across a broader segment of voters. More generally, the variable appears to capture the degree to which parties are locally representative of the communities in which they compete. Gender similarity exhibits a similar, though somewhat weaker, pattern, suggesting that representational congruence matters over and above the composition of candidate lists themselves. 

Several voter-composition variables display comparatively modest ablation effects. This does not imply that electorate demographics are unimportant. Rather, much of the information contained in municipality-level demographics is likely absorbed by the municipality embedding and historical voting patterns. Candidate characteristics and candidate--electorate similarity measures vary at the party--municipality level and therefore contribute information not already captured by broader municipal context. 

Overall, the explanatory diagnostics reveal a clear hierarchy of predictive information. Latent political structure, captured through the party and party--municipality embeddings, dominates all observed predictors. Among observed covariates, electoral persistence and organizational capacity provide the strongest signals, with the logarithm of previous vote totals, candidate-list size, and candidate--electorate age similarity emerging as the most influential variables. These findings suggest that successful municipal election forecasting relies not only on historical electoral performance but also on information describing the organizational strength and representational characteristics of local party organizations.

\begin{table}[ht]
\centering
\begin{tabular}{lr}
  \toprule
Feature & MAE change (\%) \\ 
  \midrule
Party & 8892.60 \\ 
  Interact & 625.20 \\ 
  Lag log Votes & 152.50 \\ 
  Cand n & 138.70 \\ 
  Sim Age & 90.40 \\ 
  Cand Age 30–49 & 35.30 \\ 
  Cand Age 50–64 & 28.70 \\ 
  Kommun & 26.30 \\ 
  Sim Women & 19.30 \\ 
  Lag Votes & 19.10 \\ 
  Vot First Time & 14.60 \\ 
  Vot Age 18 29 & 12.90 \\ 
  Cand Age 65+ & 11.30 \\ 
  Vot Swe Citizen & 7.30 \\ 
  Cand Women & 7.00 \\ 
  Vot Age 50 64 & 3.80 \\ 
  Vot Age 65plus & 2.90 \\ 
  Vot Age 30 49 & 2.60 \\ 
  Vot Women & 2.40 \\ 
  Has Lag Vote & 0.30 \\ 
  Cand Age Missing & 0.10 \\ 
  Has Cand & 0.00 \\ 
  Has Missing Age & 0.00 \\ 
  Vot Age Missing & 0.00 \\ 
  Cand Age 18–29 & -1.20 \\ 
   \bottomrule
\end{tabular}

\caption{ Ablation importance. \textit{Note:} Ablation importance is measured as the proportional increase in mean absolute error (MAE) after removing a feature from the fitted model,  $\Delta\mathrm{MAE} = \frac{\mathrm{MAE}^{\mathrm{ablated}}} {\mathrm{MAE}^{\mathrm{baseline}}} -1$.  Larger values indicate greater predictive importance because model accuracy deteriorates more strongly when the feature is removed. Reported values are averages across all observations. }\label{tab:ablation_elasticity}
\end{table}

\subsection{Latent Electoral Similarities Among Swedish Parties} 
\begin{figure}[!ht]
    \centering
    \includegraphics[width=0.8\linewidth]{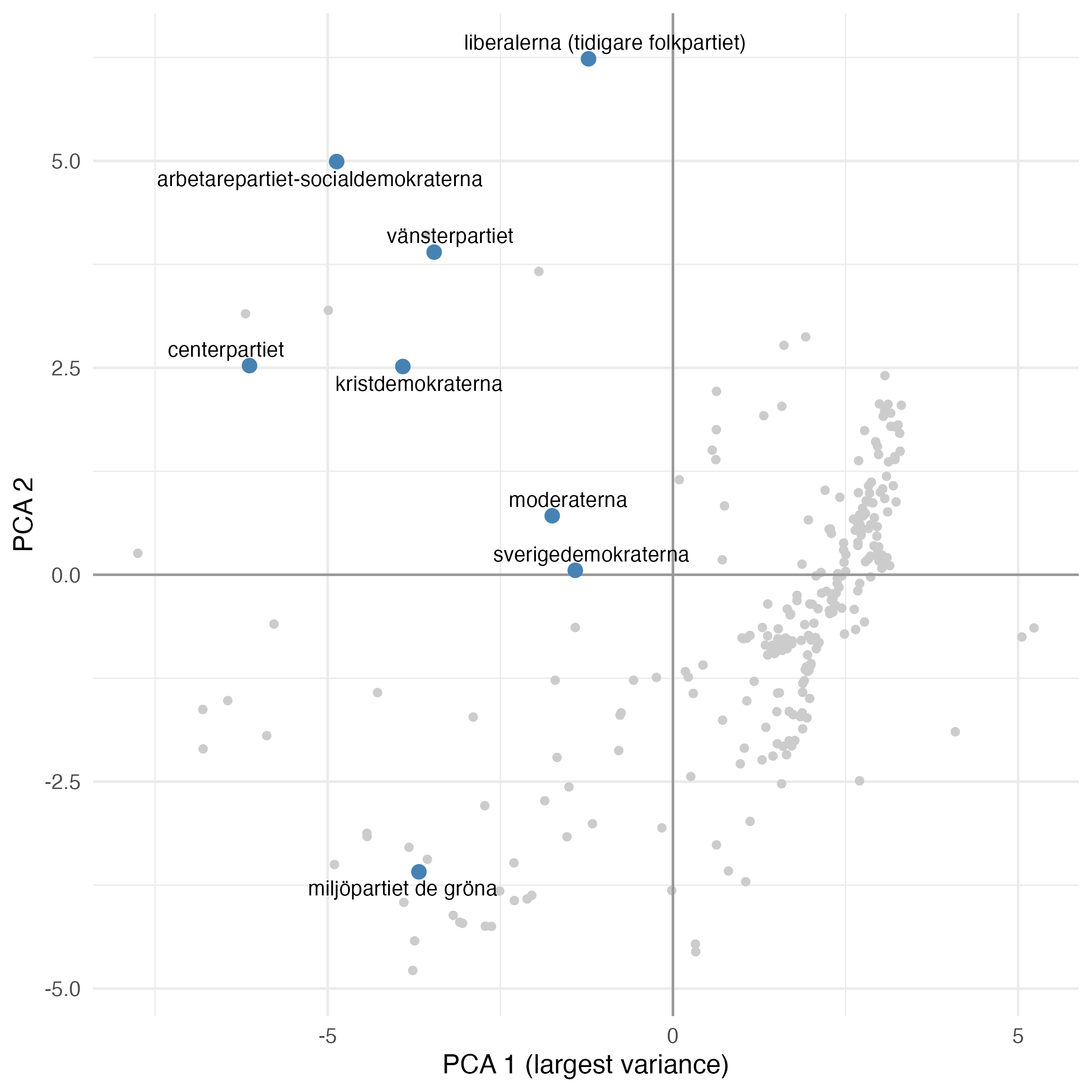}
    \caption{Latent electoral similarities among Swedish political parties. The figure shows the first two principal components of the learned party embeddings estimated by the neural zero-inflated negative binomial model. Distances in the embedding space reflect similarities in observed electoral behavior across municipalities and election cycles.}
    \label{fig:party_embedding_pca}
\end{figure}

The feature-importance analysis identified the party embedding as the  most important component of the forecasting model. Removing the party embedding increases prediction error by more than any observed covariate, indicating that persistent differences among parties contain substantial predictive information. To better understand the information encoded in this latent representation, Figure~\ref{fig:party_embedding_pca} visualizes the learned party embeddings using the first two principal components of the embedding matrix.

The principal-component representation provides a low-dimensional summary of the latent space while preserving the dominant patterns learned during model estimation. Several regularities emerge. Most notably, the major parties occupy clearly differentiated regions of the latent space rather than collapsing toward a common representation. This indicates that the model learns persistent differences in electoral performance that are not fully captured by observed demographic characteristics, candidate-list features, or historical vote measures. 

At the same time, parties with similar patterns of local electoral support tend to appear closer together. The Social Democratic Party and the Left Party occupy neighboring positions, while the Centre Party and the Christian Democrats also appear relatively close in the embedding space. Similarly, the Moderate Party and the Sweden Democrats are located in the same general region of the latent space, suggesting similarities in the geographic distribution of their support across municipalities. The Green Party and the Liberals are the two most distinctive cases. The Green Party occupies an isolated position along the negative direction of the second principal component, whereas the Liberals are separated in the opposite direction. This pattern suggests that the municipal support bases of these parties differ substantially from those of the remaining parties. 

More generally, the second principal component appears to capture a dimension of electoral differentiation that is particularly useful for separating parties with idiosyncratic geographic support patterns from the rest of the party system. It is important to emphasize that the embeddings are estimated entirely from electoral outcomes. The model is not supplied with ideological labels, coalition memberships, or any other substantive information about the parties. Consequently, Figure~\ref{fig:party_embedding_pca} should not be interpreted as a map of ideological proximity. Rather, it represents a learned measure of electoral similarity, summarizing how parties perform across municipalities and election cycles after accounting for observed demographic variables, candidate characteristics, and historical electoral performance. 

The close correspondence between the learned representation and recognizable features of Swedish municipal politics provides additional evidence that the model captures meaningful latent electoral structure. More broadly, the results illustrate the value of embedding-based approaches in electoral forecasting: instead of treating parties as unrelated categorical labels, the model learns a data-driven representation of persistent similarities and differences in local electoral competition.

\section{Forecasts for the 2026 Election}
Table~\ref{tab:nat_rank} reports the parties with the highest predicted aggregate vote shares in the 2026 municipal elections. The forecasts are obtained by first predicting party support separately within each municipality and then aggregating the resulting vote shares across municipalities. Several broad patterns emerge. First, the Social Democrats remain the dominant party in local government elections, with a predicted aggregate vote share exceeding 20\%. The Moderates constitute the second-largest party with a predicted share of approximately 15\%, while the Sweden Democrats occupy a clear third position at roughly 10\%. Together, these three parties account for a substantial share of predicted municipal support nationwide. 

Second, the model forecasts a sizable middle tier consisting of the Centre Party, the Left Party, the Christian Democrats, the Green Party, and the Liberals. Although these parties remain substantially smaller than the three largest parties, they maintain meaningful local support across Sweden and continue to play an important role in municipal politics. 

Third, the forecasts highlight the fragmented nature of local electoral competition. Beyond the eight largest parliamentary parties, a number of smaller and local parties are predicted to receive non-negligible support. These include Democraterna, Medborgerlig Samling, Partiet Nyans, Feministiskt Initiativ, and several municipality-specific electoral lists. While each of these parties accounts for only a small fraction of aggregate votes nationally, they can nonetheless become electorally relevant in particular municipalities where local political conditions favor them. The distinction between national vote share and municipal influence is particularly important in the Swedish local electoral context. Several local parties rank relatively low in aggregate predictions because they compete in only a limited number of municipalities. Nevertheless, these parties may achieve considerable support within their areas of operation and therefore remain important actors in local government formation. 

Overall, the aggregate rankings suggest broad continuity in the structure of Swedish local politics. The traditional large parties remain dominant, a second tier of established parliamentary parties retains meaningful support, and a diverse collection of smaller local parties continues to contribute to substantial variation in political competition across municipalities.

\begin{table}
\centering
\begin{tabular}[t]{lrrr}
\toprule
Party Name & Pred. votes 2026 & Pred. \% votes 2026 & Obs. \% votes 2022\\
\midrule
arbetarepartiet-socialdemokraterna & 2031054 & 29.23 \% & 29.25\%\\
moderaterna & 1489350 & 21.44 \% & 20.47\%\\
sverigedemokraterna & 949548 & 13.67 \% & 14.52\%\\
centerpartiet & 574680 & 8.27 \% & 7.67\%\\
vänsterpartiet & 530252 & 7.63 \% & 8.23\%\\
\addlinespace
kristdemokraterna & 384543 & 5.53 \% & 5.35\%\\
miljöpartiet de gröna & 370326 & 5.33 \% & 3.82\%\\
liberalerna (tidigare folkpartiet) & 296958 & 4.27 \% & 5.22\%\\
demokraterna & 38088 & 0.55 \% & 0.33\%\\
ändrings partiet revolution & 30331 & 0.44 \% & NA\\
\addlinespace
medborgerlig samling & 22887 & 0.33 \% & 0.38\%\\
kommunistiska partiet & 18584 & 0.27 \% & 0.05\%\\
partiet nyans & 16358 & 0.24 \% & 0.46\%\\
feministiskt initiativ & 12809 & 0.18 \% & 0.27\%\\
nackalistan & 8456 & 0.12 \% & 0.12\%\\
\addlinespace
piratpartiet & 7836 & 0.11 \% & 0.06\%\\
örebropartiet & 6271 & 0.09 \% & 0.12\%\\
alternativ för sverige & 5895 & 0.08 \% & 0.13\%\\
linköpingslistan & 5228 & 0.08 \% & 0.15\%\\
lidingöpartiet & 5068 & 0.07 \% & 0.08\%\\
\bottomrule
\end{tabular}
\caption{Top 20 parties by predicted aggregate vote share.  Reported percentages summarize expected national support in the 2026 municipal elections.}\label{tab:nat_rank}
\end{table}

\subsection{The Geography of Electoral Presence}\label{sec:prob_nonzero} 
As discussed in Section~\ref{sec:model}, the zero-inflation component distinguishes between party--municipality combinations that belong to the count-generating process and those that are assigned to the structural-zero state. The quantity \( 1-\pi_i \) therefore represents the probability that party \(i\) receives meaningful electoral support in a municipality rather than being effectively absent from local competition. Unlike vote shares, these probabilities measure electoral presence rather than electoral strength. 

\begin{figure}[!ht]
    \centering
    \includegraphics[width=1\linewidth]{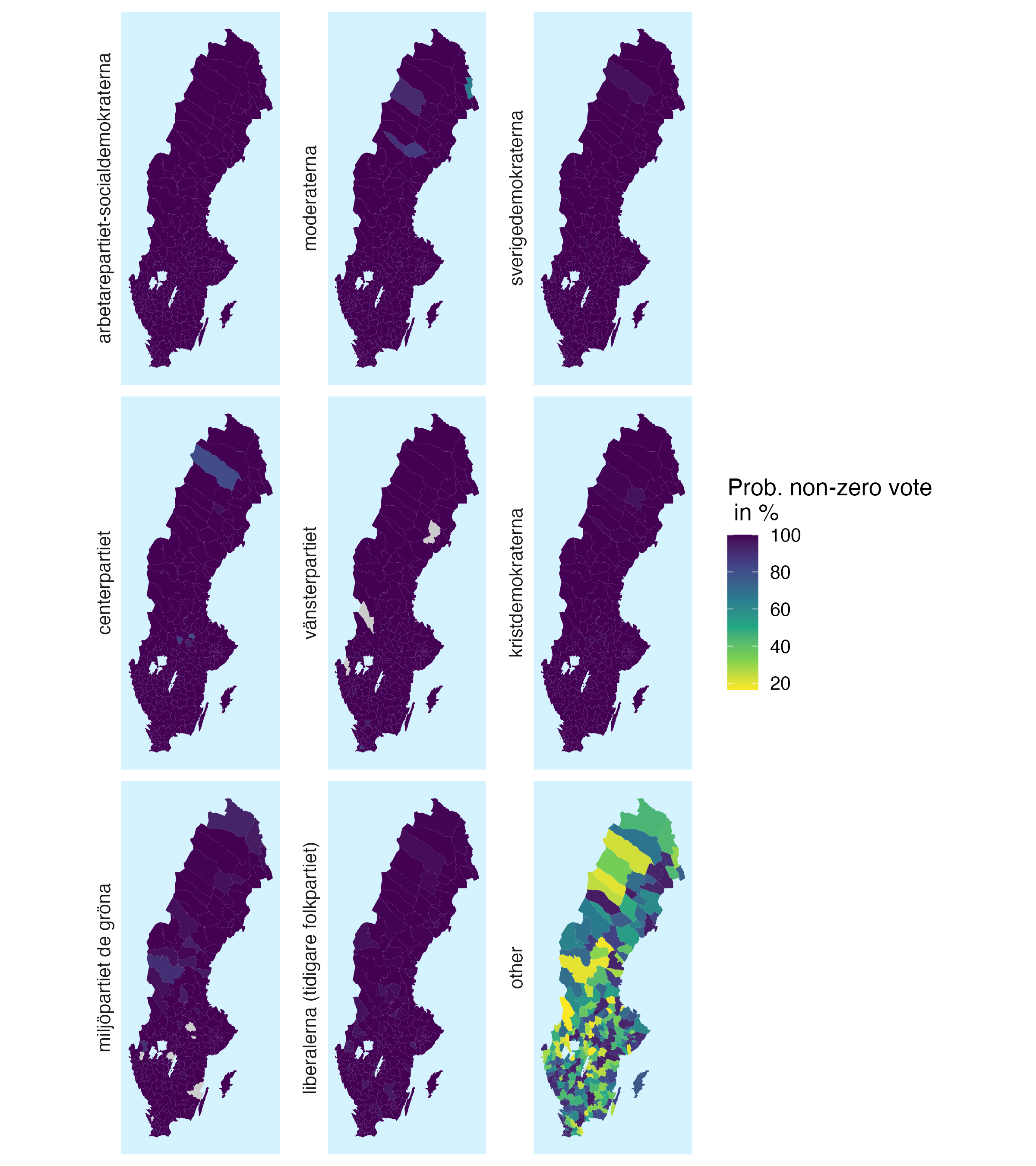}
    \caption{Predicted probability of receiving a positive vote count in the 2026 municipal election. Values are computed as $1-\hat{\pi}$ from the zero-inflation component of the model. Darker shades indicate a higher probability that a party receives at least one vote in a municipality. The established national parties are assigned near-certain electoral presence in most municipalities, whereas the residual ``other'' category exhibits substantial local variation, reflecting the geographically uneven presence of local and minor parties.} \label{fig:prob_vote}
\end{figure}

Figure~\ref{fig:prob_vote} displays the predicted probabilities of electoral presence for the major Swedish parties in the 2026 municipal election. Several broad patterns emerge. The established parliamentary parties exhibit consistently high probabilities across most municipalities, reflecting extensive organizational coverage and long-standing local electoral participation. In contrast, smaller parties and local electoral movements display more geographically concentrated patterns of presence. Their probabilities are often close to one in a limited number of municipalities while remaining substantially lower elsewhere, indicating that small parties' electoral activities are localized rather than nationally dispersed. 

The probability maps reveal an important dimension of electoral competition that is not captured by vote shares alone. A party may be expected to receive only a modest number of votes in a municipality while nevertheless showing a high probability of non-zero support, implying a stable local presence. Conversely, low values of \(1-\pi_i\) indicate municipalities in which the model considers a party unlikely to establish meaningful electoral support. The estimated probabilities therefore provide a first layer of electoral intelligence by identifying where parties are expected to compete before considering the magnitude of their support to be received. 

Altogether, the results suggest that the geography of electoral presence remains highly uneven across parties. The major parliamentary parties maintain a near-universal local footprint, whereas many smaller parties remain concentrated in specific regions or municipalities. This spatial variation in electoral presence provides the foundation for understanding the subsequent forecasts of vote counts and vote shares.

\subsection{Expected Vote Counts} \label{sec:expected_votes} 
Conditional on belonging to the count-generating process, the expected number of votes received by a party is given by the negative-binomial mean, eq.~\eqref{eq:nb_mu}. Unlike the probability maps presented in the previous subsection, which describe where parties are expected to be electorally present, the expected-count maps describe the magnitude of electoral support conditional on participation.

\begin{figure}[!ht]
    \centering
    \includegraphics[width=1\linewidth]{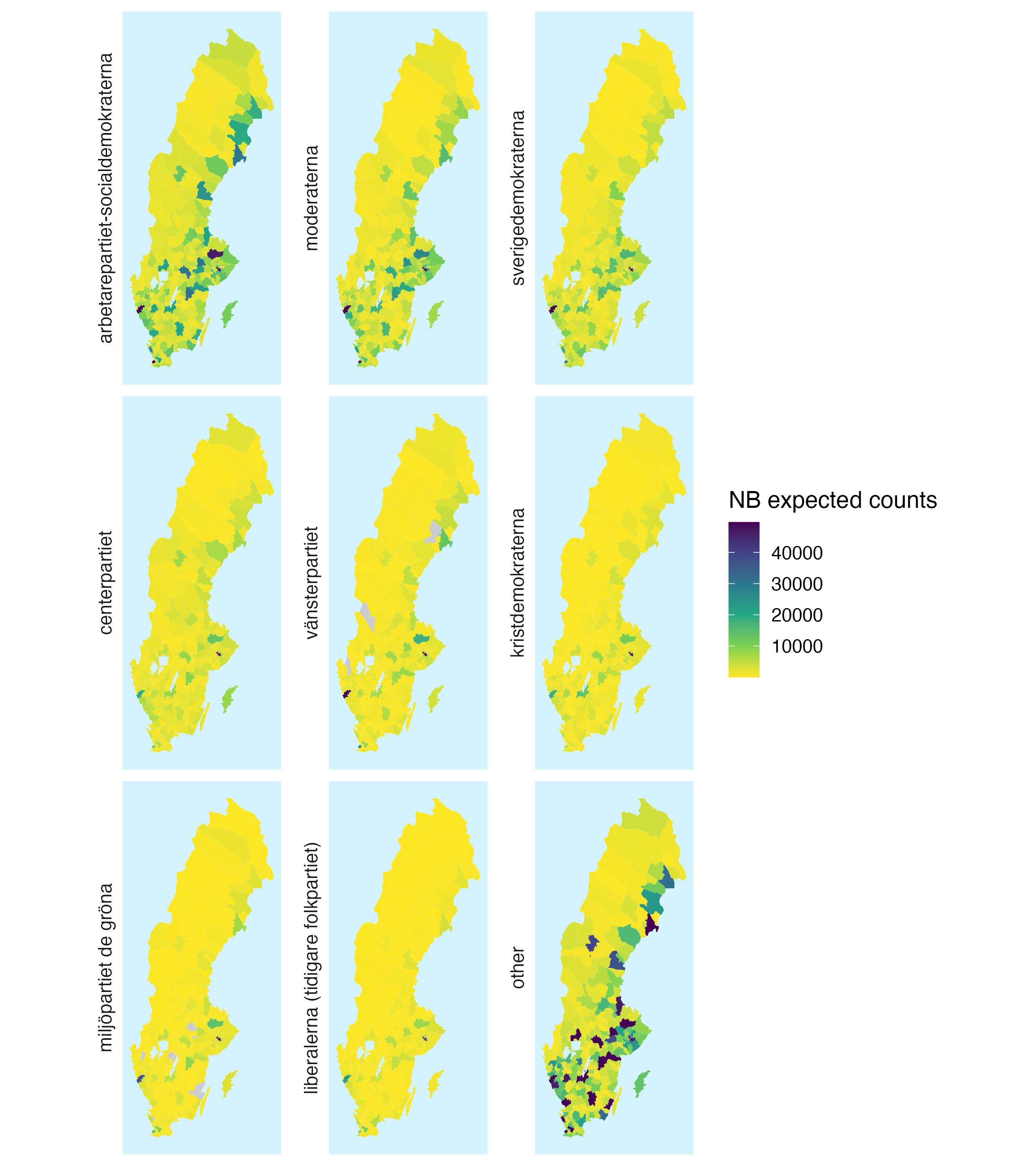}
    \caption{Predicted negative-binomial means, \(\mu_i\), for the major Swedish parties in the 2026 municipal election. Darker colors indicate larger expected numbers of votes conditional on belonging to the count-generating process. Because the maps display absolute vote counts rather than vote shares, differences reflect both party support and municipality size. } \label{fig:expected_votes}
\end{figure}

Figure~\ref{fig:expected_votes} displays the predicted values of \(\mu_i\) for the major Swedish parties in the 2026 municipal election. Several notable spatial patterns emerge. The Social Democrats are predicted to obtain the largest vote counts in many municipalities throughout the country, reflecting both their broad national presence and their continued strength in large urban areas and numerous smaller municipalities. 

The Moderates likewise exhibit substantial expected vote totals across much of Sweden, although their strongest forecasts are generally concentrated in metropolitan and suburban municipalities. 

The Sweden Democrats display a comparatively widespread electoral footprint, with expected vote counts that remain substantial across a large number of municipalities rather than being concentrated in only a few strongholds. 

The Centre Party and the Left Party exhibit more geographically uneven patterns. In particular, the Centre Party's expected support remains strongest in rural municipalities, whereas the Left Party's support is concentrated in selected municipalities where the party has historically maintained a stronger local presence. 

The Christian Democrats, Green Party, and Liberals are generally predicted to receive smaller vote totals, although each retains localized areas of relatively stronger support. 

The category ``Other'' displays the greatest degree of geographic concentration. Unlike the parliamentary parties, local parties and municipality-specific electoral lists often compete in only a small number of municipalities. As a result, expected vote counts are highly concentrated in a limited set of locations where these parties possess established organizational bases or local political relevance. Several municipalities exhibit expected vote counts for local parties that are comparable to, or even exceed, those of some parliamentary parties. 

It is important to recognize that these maps reflect absolute vote counts rather than electoral strength relative to the local electorate. Consequently, highly populated municipalities tend to exhibit larger predicted values of \(\mu_i\) simply because more votes are available. Differences in expected counts therefore combine population size with party popularity. For this reason, expected vote counts are best interpreted as forecasts of electoral scale. The next subsection normalizes these predictions by the number of eligible voters to obtain predicted vote shares, thereby facilitating comparisons across municipalities of different sizes. 

\subsection{Predicted Vote Share} 
\begin{figure}[!ht]
\centering 
\includegraphics[width=1\linewidth]{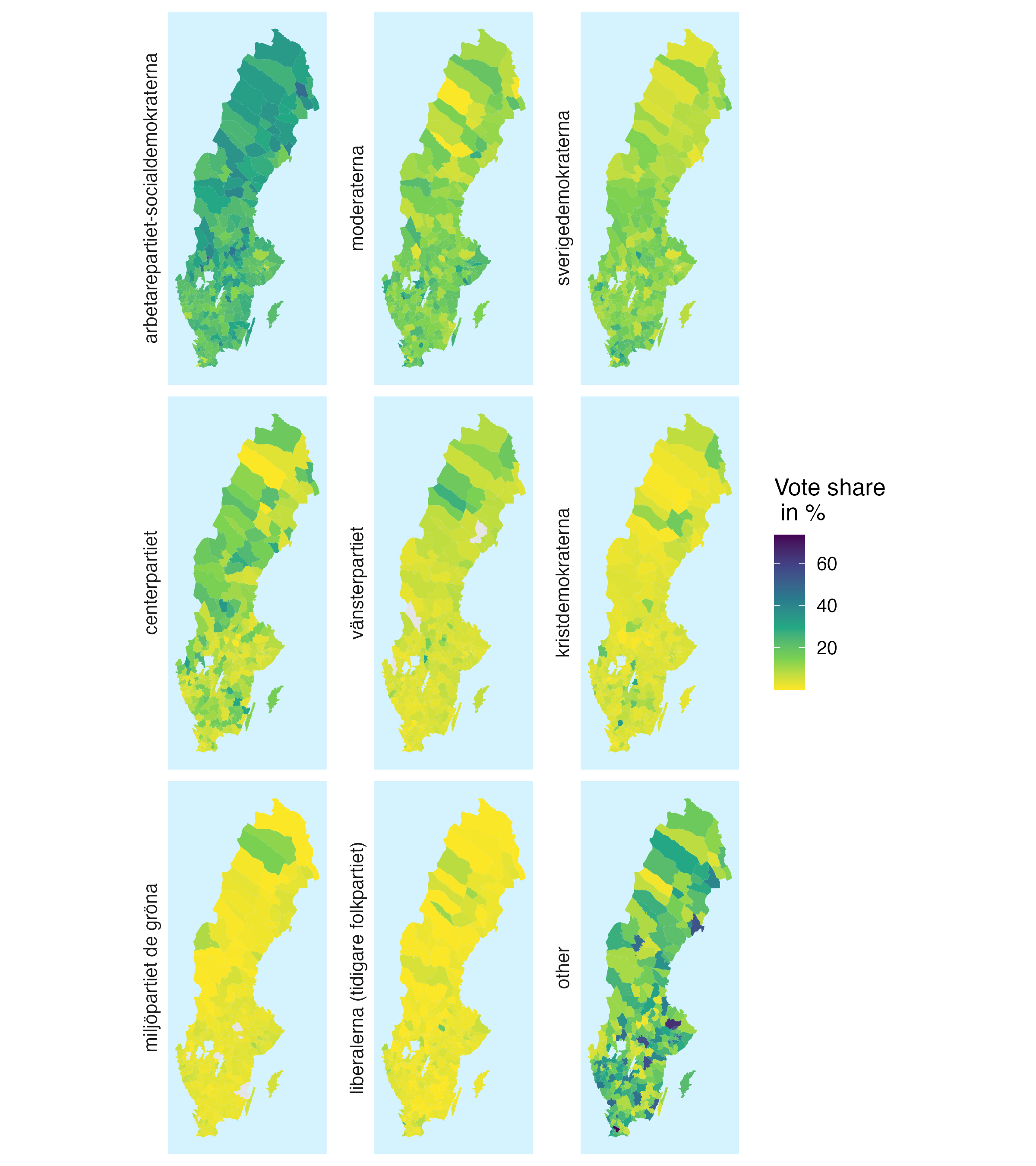} 
\caption{ Forecasted party support in the 2026 municipal elections. Colors indicate predicted vote share among eligible voters, defined as \((1-\widehat{\pi})\widehat{\mu}/N\). Darker shades indicate higher expected levels of electoral support. } \label{fig:forecast_map} 
\end{figure}

To facilitate substantive interpretation, Figure~\ref{fig:forecast_map} reports the predicted support for each party as a share of eligible voters, \[ \widehat{s}_{i} = \frac{(1-\widehat{\pi}_{i})\widehat{\mu}_{i}} {N_{mt}}, \] where \(\widehat{\pi}_{i}\) is the predicted structural-zero probability, \(\widehat{\mu}_{i}\) is the predicted negative-binomial mean, and \(N_{mt}\) is the number of eligible voters in municipality \(m\) and year \(t\). This quantity corresponds to the model's unconditional expected vote share and therefore combines information from both the count and zero-inflation components. 

The Social Democratic Party exhibits the broadest electoral footprint, with relatively strong predicted support throughout large parts of the country. Support is especially pronounced in many northern and inland municipalities. The Moderate Party are also forecasted to receive substantial support across much of Sweden, although their predicted support is generally more spatially heterogeneous than that of the Social Democrats.  The Sweden Democrats displays a different geographic profile, with stronger predicted support in many municipalities in southern and central Sweden and weaker projected support in several northern municipalities. The Centre Party shows a comparatively strong presence in many rural municipalities, consistent with its long-standing strength outside major urban centers. The Left Party and Christian Democrats are forecasted to obtain more moderate levels of support, although both parties exhibit pockets of localized strength. The Green Party and the Liberals display the lowest predicted vote shares among the established national parties. Their support is concentrated in a relatively small number of municipalities, producing a more fragmented spatial pattern than those observed for the larger parties. The residual ``other'' category exhibits the greatest spatial heterogeneity of all groups, reflecting the uneven geographic distribution of local parties and municipality-specific electoral lists. 

While support for the established national parties varies relatively smoothly across space, support for local and residual parties is much more concentrated and municipality-specific. Overall, the forecasts suggest that municipal electoral competition in 2026 is characterized by a combination of broad national party presence and substantial local heterogeneity. National parties dominate most municipalities, but local electoral environments continue to generate meaningful variation in party support across the country.

\subsection{Party Ranks by Vote Share within Municipalities}
\begin{figure}
    \centering
    \includegraphics[width=1\linewidth]{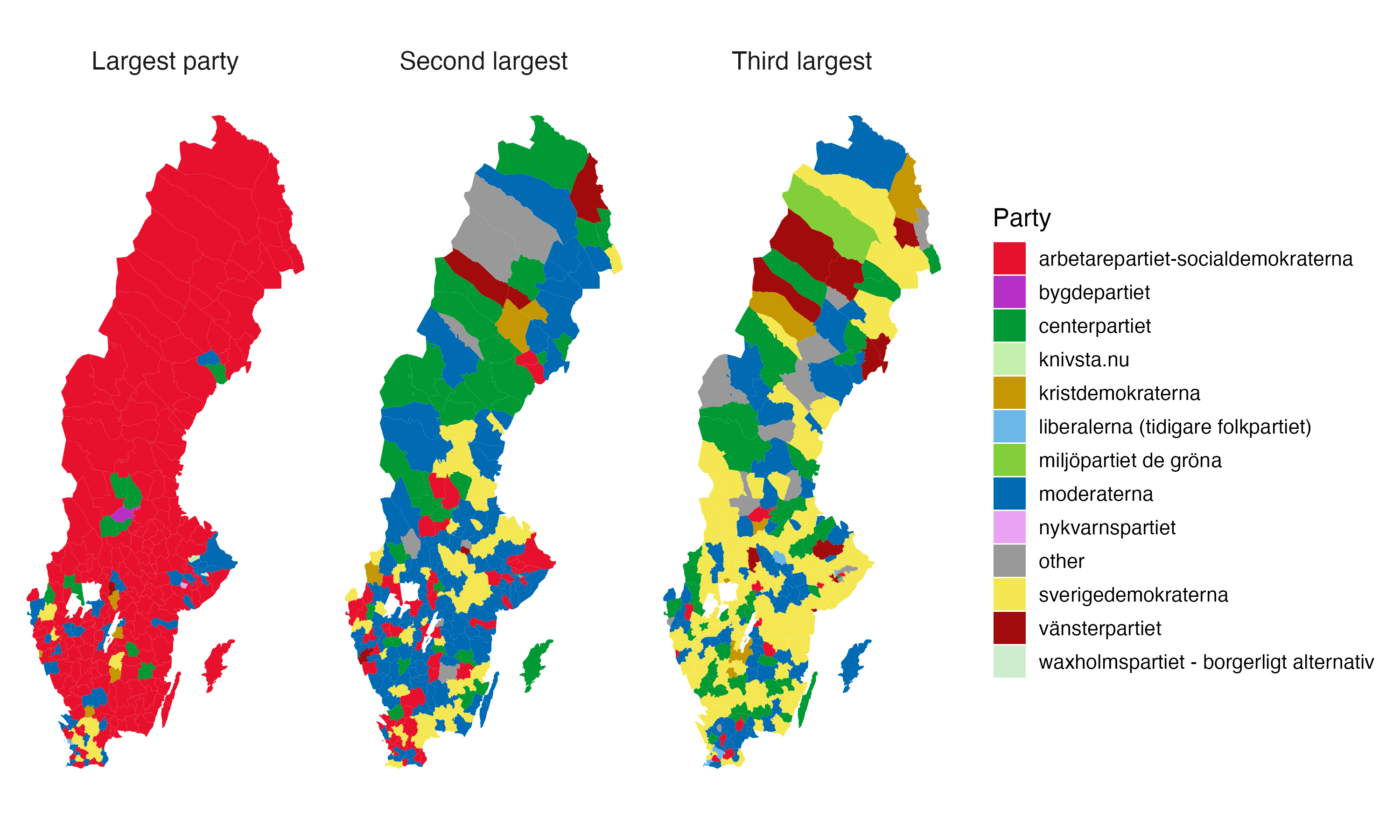}
    \caption{ Predicted party rankings within municipalities in the 2026 municipal elections. Each panel identifies the party with the largest, second-largest, and third-largest forecasted vote share among eligible voters. } \label{fig:forecast_rank}
\end{figure}

While Figure~\ref{fig:forecast_map} illustrates the intensity of expected party support, it does not directly reveal the competitive ordering of parties within municipalities. Figure~\ref{fig:forecast_rank} therefore summarizes the predicted rank ordering of parties according to forecasted vote shares. For each municipality, the figure identifies the party expected to finish first, second, and third based on eq.~\eqref{eq:pred_share}.

Several patterns become considerably clearer once party rankings are examined directly. Most notably, the Social Democratic Party is forecasted to remain the largest party in a substantial majority of municipalities. The party dominates much of northern Sweden and large portions of central Sweden, reflecting its broad national presence and historically strong position in municipal politics. However, this traditional dominance is losing ground in the south part of the country, especially along the southwest coast and in Skane where the Moderates and Sweden Democrates constitutes the principal challenger to the Social Democrats. 

Although the Moderates are forecasted to be the largest party in relatively few municipalities, they frequently appear as the second-largest competitor and are particularly visible in southern and metropolitan areas. The Sweden Democrats display a similar pattern. Rather than consistently occupying first place, they are most often forecasted to rank second or third, indicating widespread electoral strength but fewer municipalities in which they are expected to become the dominant party. 

The second- and third-place maps reveal substantially greater geographic fragmentation than the map of leading parties. Several parties emerge as regionally important competitors despite rarely being forecasted to finish first. The Centre Party occupies prominent positions in many rural municipalities, especially in central and northern Sweden. The Christian Democrats, Left Party, Green Party, and Liberals also appear in localized clusters where they are forecasted to remain relevant competitors even when not leading municipal contests. 

Local parties and municipality-specific electoral lists play a limited role among the largest-party forecasts but appear more frequently in the lower ranking categories. This pattern is consistent with the vote-share maps presented earlier: local parties rarely dominate municipal elections nationally, yet they remain important competitors in specific municipalities where local political issues and organizational structures create distinct electoral environments. 

Taken together, the ranking analysis suggests that the forecasted 2026 municipal elections are characterized by a combination of national dominance and local variation. The Social Democrats are expected to remain the leading force across most municipalities, while competition for second and third place is considerably more heterogeneous and geographically differentiated. This finding highlights that much of the variation in Swedish municipal politics arises not from changes in the dominant party, but from differences in the configuration of competing parties beneath the leading position.

\subsection{Forecast Uncertainty} \label{sec:forecast_uncertainty} 
\begin{figure}[!ht]
    \centering
    \includegraphics[width=1\linewidth]{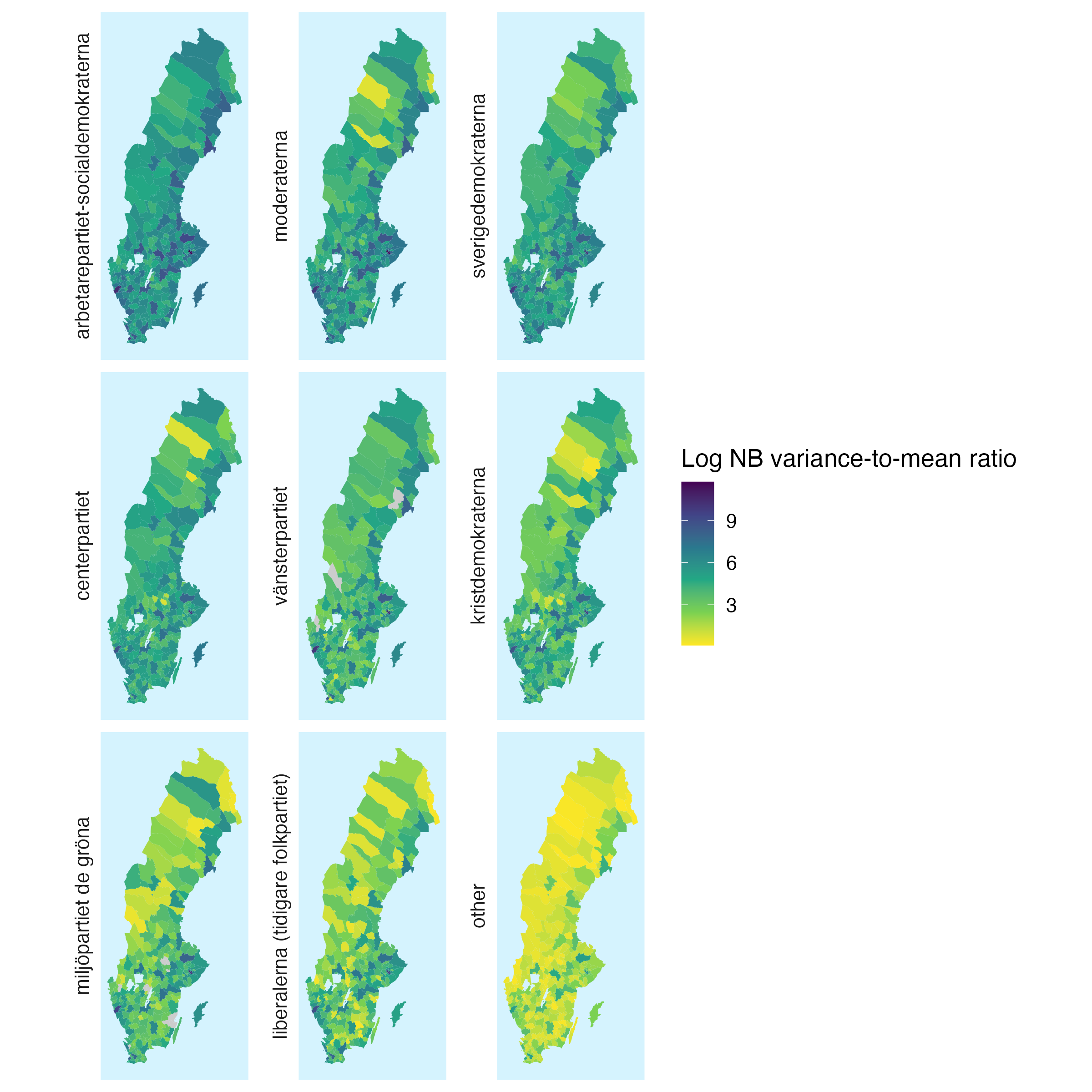}
    \caption{ Predicted uncertainty in party vote counts for the 2026 municipal election, measured as the logarithm of the negative-binomial variance-to-mean ratio, \( \log\{\mathrm{Var}(y_i\mid\mu_i,\alpha_i)/\mu_i\} \). Higher values indicate greater forecast uncertainty relative to the expected number of votes. } 
    \label{fig:forecast_var_mu}
\end{figure}

The expected vote counts presented in the previous subsection describe the center of the predictive distribution. However, electoral forecasting also requires an assessment of uncertainty around those expectations. For a negative-binomial outcome, uncertainty is governed by both the conditional mean and the municipality-specific dispersion parameter: \[ \mathrm{Var}(y_i \mid \mu_i,\alpha_i) = \mu_i+\alpha_i\mu_i^2. \] Because the variance naturally increases with the expected number of votes, direct maps of \(\mathrm{Var}(y_i\mid\mu_i,\alpha_i)\) are difficult to compare across municipalities and parties. To isolate uncertainty relative to the predicted mean, Figure~\ref{fig:forecast_var_mu} therefore reports the logarithm of the variance-to-mean ratio, \[ \log\left( \frac{ \mathrm{Var}(y_i\mid\mu_i,\alpha_i) }{ \mu_i } \right), \] which summarizes the degree of over-dispersion implied by the model. 

Several patterns emerge from the maps. First, uncertainty varies systematically across both municipalities and parties. Highly populated municipalities in southern and central Sweden frequently exhibit larger variance-to-mean ratios than many rural municipalities, indicating that electoral outcomes in these areas remain more difficult to predict even after conditioning on electoral history, candidate characteristics, electorate composition, and latent party structure. These municipalities often contain more complex political environments with greater party competition and larger voter populations, allowing for a wider range of plausible electoral outcomes. 

Second, the degree of forecast uncertainty differs substantially across parties. The largest parties, including the Social Democrats, Moderates, and Sweden Democrats, display high uncertainty in many municipalities despite having relatively stable expected vote counts. This pattern reflects the fact that larger expected vote totals generate greater absolute variability even when electoral support is comparatively predictable. By contrast, parties with smaller expected vote counts often exhibit lower variance-to-mean ratios, particularly in municipalities where their electoral presence is weak and support remains consistently limited. 

Third, the category ``Other'' displays noticeably lower uncertainty than many parliamentary parties despite containing numerous smaller parties and local electoral lists. This result reflects the localized nature of support for many minor parties. In municipalities where these parties have little electoral presence, both expected vote counts and associated uncertainty remain low. When strong local parties are present, uncertainty increases, but these municipalities remain relatively concentrated geographically. 

More broadly, the uncertainty maps reveal a dimension of electoral intelligence that is not visible from expected vote counts or predicted vote shares alone. Two municipalities may exhibit similar expected support for a party while having substantially different levels of forecast uncertainty. Likewise, a party may be expected to perform strongly in a municipality while still facing considerable uncertainty regarding the exact magnitude of that support. The variance-to-mean ratio therefore provides a measure of forecast reliability, complementing the point forecasts reported in previous sections. 

Taken together, the results suggest that electoral uncertainty is itself spatially structured. The model predicts not only where parties are expected to perform well, but also where those forecasts should be interpreted with greater caution. Consequently, the forecasts presented here should be understood as probabilistic predictions whose reliability varies across municipalities and parties rather than as deterministic projections of future electoral outcomes.

\section{Conclusion} 
Forecasting local elections presents challenges that differ substantially from those encountered in national electoral forecasting. Municipal elections involve multiparty competition, strong geographic heterogeneity, frequent party entry and exit, and a large number of party--municipality combinations receiving little or no electoral support. These characteristics generate count outcomes that are heavily over-dispersed and contain many excess zeros, making conventional forecasting approaches less suitable. This paper developed a bounded neural zero-inflated negative binomial framework designed specifically to address these features of local electoral data. 

The proposed approach combines several elements. A zero-inflated negative binomial likelihood captures excess zeros and over-dispersion, while a bounded vote-share formulation ensures that expected vote counts remain consistent with the size of the eligible electorate. Party, municipality, and party--municipality embeddings allow the model to learn persistent political structure across municipalities, and ablation analysis provide interpretability. Municipality-specific dispersion parameters further enable uncertainty to vary across local electoral environments. 

The empirical results reveal a clear hierarchy of predictive information. Party identity and party--municipality interactions dominate prediction, indicating that local electoral competition is strongly shaped by persistent partisan and geographic structures. Among observed predictors, electoral history remains the most important source of information. Candidate-list characteristics and candidate--electorate similarity also contribute substantially to predictive performance, whereas municipality-level demographic composition generally plays a secondary role. Particularly noteworthy is the finding that demographic similarity between candidates and electorates appears more informative than the demographic composition of candidates alone, suggesting that representational congruence may play an important role in local electoral competition. 

The forecasts for the 2026 municipal elections suggest broad continuity in the Swedish party system. The Social Democrats remain the dominant local political force. However, this dominance is likely to lose ground in the southern part of Sweden, particulalry in Skane, to the Moderates and Sweden Democrats. Meanwhile, the remaining parliamentary parties retain meaningful but more geographically concentrated support. The forecasts also highlight the continuing importance of local political movements, many of which remain electorally relevant despite limited national visibility. Beyond point forecasts, the model identifies substantial variation in the geographic distribution of electoral presence and forecast uncertainty. Consequently, the analysis provides information not only about expected electoral outcomes but also about where those forecasts should be interpreted with greater caution. 

Several limitations should be acknowledged. First, the 2026 forecasts rely on the most recent available electorate-composition information and cannot account for demographic changes occurring after data collection. Second, while the neural architecture captures complex nonlinear relationships, the resulting predictions remain observational and should not be interpreted as causal effects. These limitations suggest several directions for future research. One possibility is to incorporate dynamic demographic projections, polling information, or campaign indicators as they become available during the electoral cycle. More generally, the framework introduced here demonstrates how modern machine-learning methods can be integrated with interpretable count-data models to address challenging forecasting problems in political science. As increasingly detailed electoral and administrative data become available, such approaches may provide a useful foundation for forecasting local elections in Sweden and in other multiparty democracies.

\section*{Acknowledgments}
Financial support is gratefully noted from the Swedish Research Council (Vetenskapsrådet) grant \texttt{2022-06012\_3} (Climate-Induced Migration in Africa and Beyond -- CLIMB) and grant \texttt{2023-06147} (New perspectives on the electoral participation of immigrants in Sweden).

\section*{Data and Code Availability} 
All code used for model estimation, forecasting, and figure generation is publicly available at \url{https://github.com/HaodongQi/voting_swe_2026}. Raw electoral and candidate data are obtained from official Swedish public sources and can be downloaded from \url{https://www.val.se/valresultat-och-statistik/statistik-och-data/radata-fran-val-2002-2022}. Forecasts reported in this paper were publicly released prior to the election through a GitHub repository \url{https://github.com/HaodongQi/voting_swe_2026/releases/tag/v1.0}. 

\bibliographystyle{unsrt}  
\bibliography{references}

\end{document}